# High-Fidelity ROI CT Reconstruction with Limited Quantum Resources via Hybrid Classical-Quantum Refinement

Hyunju Lee
*Department of mathematics*
*Dongguk University*
Seoul, South Korea.
*Quantum Research Center*
*QTomo Inc.*
Cheongju, South Korea.
Email: hjhj2627@dongguk.edu

Jeonghwa Lee
*Quantum Research Center*
*QTomo Inc.*
Cheongju, South Korea..
Email: jasminhlee@gmail.com

Kyungtaek Jun*
*Quantum Research Center*
*QTomo Inc.*
Cheongju, South Korea.
Email: ktfriends@gmail.com

***Abstract*— Quantum optimization for computed tomography (CT) reconstruction is constrained by the number of binary variables required for image representation, making direct whole-image quantum reconstruction difficult for large or structurally complex objects. We propose a hybrid region-of-interest (ROI) refinement framework in which a coarse global image is first reconstructed by quantum tomographic reconstruction (QTR) and quantum compressed sensing tomographic reconstruction (QCSTR), filtered backprojection (FBP), or simultaneous algebraic reconstruction technique (SART), and quantum optimization is then applied only to the selected ROI through a residual projection-image formulation. This strategy reduces the effective QUBO size while preserving high-fidelity reconstruction in the target region. Experiments on three discrete phantom samples show that both QTR/QCSTR+QTR/QCSTR and SART+QTR/QCSTR achieve accurate ROI reconstruction for moderate-size cases under a reduced-angle setting. For the largest and most complex sample, the quality of the coarse global estimate becomes critical, and the best result is obtained when a stable classical coarse reconstruction is combined with second-stage ROI-only QTR/QCSTR. Among the tested pipelines, SART+QTR/QCSTR achieves the lowest average ROI RMSE and MAE. The results indicate that the practical advantage of quantum-assisted CT reconstruction lies in reserving quantum optimization for local refinement while using classical reconstruction to stabilize the global background.**

*Corresponding author: Kyungtaek Jun.

## I. Introduction

Computed tomography (CT) is a fundamental imaging modality in medicine, biology, and industrial inspection because it enables non-destructive visualization of internal structures from projection measurements [1], [2]. In many practical scenarios, however, the entire image does not need to be reconstructed at uniformly high quality. Instead, diagnostic or engineering decisions often depend on a limited region of interest (ROI), such as the heart, a suspected lesion, or a localized defect [1], [3]. This makes ROI-focused reconstruction particularly important in settings where measurement data, radiation dose, or computational resources are limited [1], [3].

A major challenge arises when CT reconstruction must be performed from reduced-angle or sparse-view measurements. In such settings, the inverse problem becomes increasingly ill-posed, and classical reconstruction methods often suffer from artifacts or loss of fine local detail [2], [4]. Filtered backprojection (FBP) remains attractive because of its simplicity and speed, but it is well known to produce pronounced streaking and missing-edge artifacts when projection angles are insufficient [2], [4]. Iterative methods such as simultaneous algebraic reconstruction technique (SART) are generally more robust under incomplete-data conditions, yet they still do not fully overcome the degradation of local structural accuracy under strong undersampling [4], [13]. These limitations are especially critical when the goal is not merely to produce a plausible global image, but to improve reconstruction fidelity in a selected local region.

Recent studies have shown that quantum optimization provides a promising alternative formulation for tomographic imaging by expressing CT reconstruction as a quadratic unconstrained binary optimization (QUBO) problem [5]–[8]. In this framework, the image is discretely represented by binary variables, and reconstruction is performed by minimizing the mismatch between the measured projection image and the projection image generated from the qubit-encoded image [5]. This line of work has evolved from initial full-image QUBO reconstruction to more stable pixel encodings, segmentation-oriented formulations, compressed-sensing regularization, and limited-qubit refinement strategies [5]–[8]. However, these studies also make clear that direct whole-image quantum reconstruction becomes increasingly difficult as image size grows, because the number of variables and the conditioning of the QUBO problem both become challenging under current hardware and solver limitations [5]–[8].

More broadly, the search for quantum advantage has motivated both benchmark-oriented and application-oriented studies. While strong forms of quantum advantage remain highly problem-dependent, recent work has increasingly focused on whether quantum methods can provide meaningful practical benefit in specific scientific and engineering tasks. Evidence of scaling advantage has been reported for selected optimization benchmarks, and hybrid quantum-centric workflows have also demonstrated useful gains in chemistry and materials-design applications [9]–[11]. In this context, our previous QTR/QCSTR studies are better viewed as demonstrating problem-specific quantum benefit in constrained CT reconstruction settings: they show that quantum optimization can remain competitive in standard regimes and can succeed in extreme scenarios where classical reconstruction becomes unreliable [5]–[8].

Motivated by this limitation, we investigate a different strategy in which quantum resources are reserved only for the most important local region rather than for the entire high-resolution image. Specifically, we study an ROI-focused hybrid framework for limited-angle CT reconstruction in which a coarse global image is first obtained by QTR/QCSTR, FBP, or SART, and quantum optimization is then applied only to the selected ROI through a residual projection-image formulation [7]. In this setting, the coarse reconstruction provides a coarse global estimate, while the target region is refined by solving a second QUBO problem based on the residual projection image. This design keeps the effective quantum optimization problem small while still targeting high-fidelity recovery where it is most needed [7].

The central question addressed in this paper is therefore not whether the full CT image should be reconstructed quantum mechanically, but how classical and quantum stages should be combined most effectively under limited quantum resources. To answer this question, we compare a fully quantum two-stage pipeline with hybrid strategies that use classical coarse reconstruction followed by QTR/QCSTR-based ROI refinement. In this way, the present study aims to clarify the practical role of quantum optimization in CT reconstruction, namely as a tool for targeted local refinement built on top of a reliable coarse estimate rather than as a direct replacement for full-image reconstruction.

Because the number of variables required by QTR/QCSTR grows rapidly with the size of the reconstruction domain, applying fully quantum CT reconstruction to large-scale images remains difficult under currently available hardware constraints [5]–[8]. For this reason, the present study focuses on whether QTR/QCSTR can still provide meaningful value when restricted to clinically or practically important local regions. We therefore position this work as a proof of feasibility for high-fidelity localized reconstruction under limited quantum resources. In medicine, such a strategy may be relevant to future precision-focused imaging workflows in which only a diagnostically important local structure is reconstructed with higher fidelity, while the surrounding background is represented more coarsely. More generally, the same idea is also applicable to industrial or scientific imaging tasks in which the target of interest occupies only a limited portion of the full field of view. In this sense, the present work suggests the potential of ROI-prioritized imaging strategies in which reconstruction fidelity is concentrated on the most important local structure rather than distributed uniformly over the entire image.

## II. Background and Related Work

### A. CT Reconstruction Under Reduced-Angle Measurements

In X-ray CT, the measured projection data are represented as a sinogram, which can be mathematically modeled by the Radon transform [2], [12]. For an image $f(x, y)$, the Radon transform is defined as

$$\mathcal{R}f(\theta, s) = \int_{-\infty}^{\infty} f(s\cos\theta - t\sin\theta, s\sin\theta + t\cos\theta)dt, \quad (1)$$

where $\theta$ denotes the projection angle and $s$ denotes the detector position. This expression represents the line integral of the image intensity along the line $x\cos\theta + y\sin\theta = s$. Accordingly, CT reconstruction is formulated as the inverse problem of recovering an image from its sinogram [2], [12].

In practice, projection measurements are acquired at multiple angles, and reconstruction quality strongly depends on the amount and coverage of these measurements [2], [4]. Classical reconstruction methods such as filtered backprojection (FBP) and simultaneous algebraic reconstruction technique (SART) are widely used to approximate the inverse Radon transform [2], [13]. FBP is computationally efficient and has long served as a standard reconstruction approach in CT, whereas SART is an iterative method that is generally more robust when the available measurements are incomplete or noisy [2], [13]. However, when the number of projection angles is reduced, the reconstruction problem becomes more ill-posed, and both methods suffer from degraded performance [2], [4]. In particular, FBP tends to produce pronounced streaking artifacts and missing-edge structures, while SART, although typically smoother and more stable, still cannot fully recover fine local details under strongly undersampled conditions [2], [4]. This reduced-angle setting is directly relevant to the present study, since the goal is not to reconstruct the entire image uniformly, but to improve reconstruction fidelity in a selected region of interest under limited measurement information.

### B. QUBO-Based Quantum CT reconstruction

Quantum optimization offers an alternative way to formulate CT reconstruction as a discrete optimization problem [5]. In QUBO-based approaches, pixel values are encoded by binary variables, and the Radon transform of the encoded image yields a qubit-encoded sinogram [5], [6]. The reconstruction problem is then written as the minimization of the discrepancy between this encoded projection image and the experimentally measured projection image, resulting in a quadratic unconstrained binary optimization problem that can be solved by quantum annealers or hybrid quantum-classical solvers [5], [6]. In this way, the continuous inverse problem of CT reconstruction is recast as a combinatorial optimization problem over binary variables.

The earliest QUBO-based CT reconstruction studies demonstrated that accurate image recovery could be achieved from sinogram patterns by directly optimizing a qubit-encoded image [5]. Subsequent work extended the same principle to segmentation directly from X-ray data, showing that physically meaningful discrete encodings can also be effective in tomographic analysis [6]. These studies established the feasibility of quantum-assisted CT imaging and motivated later efforts to improve scalability and robustness.

### C. From Whole-Image Quantum Reconstruction to ROI-Only Residual Refinement

Although QUBO-based CT reconstruction has shown promising results, direct whole-image quantum reconstruction becomes increasingly difficult as image size increases [5], [7], [8]. Early binary encoding strategies often required many variables and produced QUBO coefficients with a large dynamic range, which reduced solver stability on larger problems [5]. To address this limitation, later studies introduced range-dependent formulations, physically meaningful mass attenuation coefficient (MAC)-difference encodings, compressed-sensing regularization, and limited-qubit refinement strategies [6]–[8]. These developments significantly improved practical performance, but they also highlighted a persistent limitation: full high-resolution quantum reconstruction remains difficult under current qubit and solver constraints [7], [8].

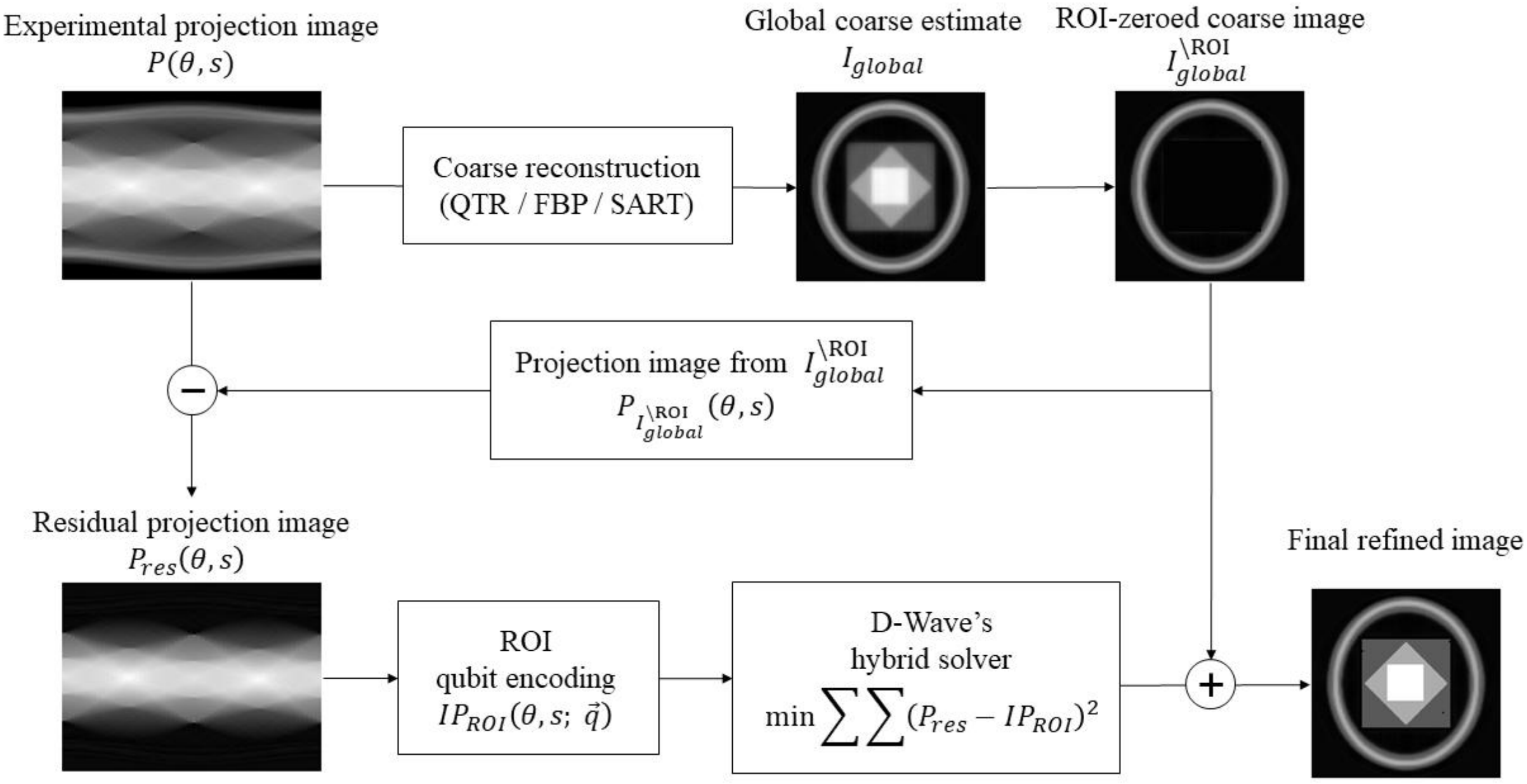

Fig. 1. Overview of the proposed hybrid ROI-refinement pipeline for Sample 1 using the SART+QTR configuration. SART first produces the global coarse estimate $\boldsymbol{I_{global}}$. After zero-masking the ROI in $\boldsymbol{I_{global}}$, its projection image is subtracted from the experimental projection image $\boldsymbol{P(\theta, s)}$ to form the residual projection image $\boldsymbol{P_{res}(\theta, s)}$. The ROI is then refined by solving a QUBO with the D-Wave hybrid solver, and the optimized ROI is inserted back into the coarse image to obtain the final refined reconstruction.

The present work is motivated by this limitation and shifts the focus from whole-image recovery to local refinement [7]. Instead of encoding the entire high-resolution image into a single large QUBO problem, we first construct a coarse reconstruction using QTR/QCSTR, FBP, or SART, and then encode only the selected ROI into binary variables. A residual projection image is formed by subtracting the coarse background contribution from the measured projection image, and the resulting QUBO is used only for local refinement. This ROI-only residual refinement framework reduces the effective size of the quantum optimization problem and allows quantum resources to be concentrated on the most important part of the image [7]. From this perspective, the contribution of quantum optimization is not necessarily to replace classical reconstruction completely, but to complement it by providing a targeted refinement mechanism where classical coarse reconstruction alone is insufficient.

Accordingly, the main methodological question considered in this paper is how coarse global reconstruction and local quantum refinement should be combined under limited quantum resources. The proposed framework for addressing this question is presented in the next section.

## III. Method

This section describes the proposed reconstruction framework, including phantom generation, coarse global reconstruction, residual projection-image formation, QUBO-based ROI refinement, and the evaluation metrics used in the experiments. An overview of the full reconstruction pipeline is shown in Fig. 1.

### A. Problem Setup

We consider a CT reconstruction problem in which the measured projection data are given as a reduced-angle sinogram projection image $P$. Let the target image be defined on an $N \times N$ grid, and let $\Omega_{ROI}$ denote the selected region of interest (ROI) to be refined at high fidelity. The objective of the proposed method is not to reconstruct the full image entirely through quantum optimization, but to obtain a coarse global estimate first and then refine only the ROI through a QUBO-based quantum optimization step.

In the proposed framework, the coarse global reconstruction is obtained by one of three methods: QTR/QCSTR, FBP, or SART. This coarse image is then used to construct a coarse global estimate at the target resolution. The ROI is subsequently modeled using binary decision variables through a difference-MAC-based discrete representation, and a residual projection image is formed so that the quantum optimization step focuses only on the remaining discrepancy associated with the selected region.

### B. Phantom Design and Measurement Setup

The synthetic phantom samples were constructed for evaluation. Samples 1 and 2 were defined on $120 \times 120$ grids, whereas Sample 3 was defined on a $180 \times 180$ grid. In all three cases, the target region of interest (ROI) had size $60 \times 60$. The phantoms were manually designed to contain discrete intensity levels and structurally distinct local patterns so that the effectiveness of ROI refinement could be examined under different geometric conditions. In particular, Samples 1 and 2 were used to study ROI refinement in moderate-size images, whereas Sample 3 was used to test the same refinement strategy in a larger image with the same ROI size.

For all samples, projection data were generated numerically by applying the Radon transform to the phantom image. To create a consistent reduced-angle reconstruction setting, all sinograms were generated using 60 projection angles uniformly sampled over $[0°, 180°)$, regardless of image size. Thus, even though Samples 1 and 2 have size $120 \times 120$ and Sample 3 has size $180 \times 180$, all three cases were reconstructed from the same number of projection views. In this paper, the term reduced-angle refers to this fixed 60-view acquisition setting, which is more restrictive than the commonly used case in which the number of projection angles scales with image size.

The resulting sinograms therefore have detector resolution determined by the image size, while the number of projection views is fixed at 60. This setup allows us to compare the proposed method across phantoms of different sizes under a common measurement budget and to examine how effectively the ROI can be refined when the available angular information is limited.

*C. Coarse reconstruction and background modeling*

Let $I_{coarse}$ denote the coarse reconstruction obtained from the reduced-angle sinogram. In this study, the coarse reconstruction is generated by one of three methods: QTR/QCSTR, FBP, or SART. The purpose of this stage is not to produce the final high-fidelity reconstruction, but to obtain a global approximation of the image before ROI refinement.

When QTR/QCSTR is used, reconstruction is first performed on a reduced-resolution grid in order to keep the quantum optimization problem within the available qubit budget. The resulting image is then resized to the target resolution and mildly smoothed with a Gaussian filter to suppress staircase artifacts introduced by upsampling. When FBP or SART is used, the coarse reconstruction is obtained directly at the target resolution and used without additional smoothing.

We denote the resulting global coarse estimate by $I_{global}$, which is defined as

$$I_{global} = \begin{cases} \mathcal{G}_\sigma(\mathcal{U}(I_{coarse})), & \text{for } I_{coarse} \text{ from QTR/QCSTR,} \\ I_{coarse}, & \text{for } I_{coarse} \text{ from FBP or SART.} \end{cases} \quad (2)$$

This image is not treated as the final reconstruction. Instead, it is used to explain the portion of the measured projection image that can already be accounted for by the coarse global reconstruction before local quantum refinement is applied. In this sense, the coarse image serves as a global coarse estimate for residual projection-image formation rather than as the final solution itself.

*D. Residual projection image and ROI qubit representation*

Since only the selected ROI is refined in the second-stage QUBO optimization, we define $I_{global}^{\setminus ROI}$ as the image obtained by setting the ROI pixels in $I_{global}$ to zero. This image preserves the coarse full-image estimate outside the ROI while removing the contribution of the target region.

Let $P(\theta, s)$ denote the experimental projection image obtained from the measured sinogram. We further define

$$P_{I_{global}^{\setminus ROI}}(\theta, s)$$

as the projection image generated from $I_{global}^{\setminus ROI}$. The residual projection image is then given by

$$P_{res}(\theta, s) = P(\theta, s) - P_{I_{global}^{\setminus ROI}}(\theta, s). \quad (3)$$

To restrict quantum optimization to the target region, only the ROI pixels are represented by binary decision variables. Let $I^{ij}$ denote the pixel value at location $(i, j)$ in the ROI image. Following our previous MAC-based formulation, each ROI pixel is expressed using increasing attenuation levels $\alpha_1 < \alpha_2 < \cdots < \alpha_K$ as

$$I^{ij} = \alpha_1 q_1^{ij} + \sum_{k=2}^{K} (\alpha_k - \alpha_{k-1}) q_k^{ij}, \quad q_k^{ij} \in \{0, 1\}. \quad (4)$$

This difference-MAC-based representation provides a physically meaningful discrete encoding while avoiding the instability associated with radix-style binary expansions. Using this representation, we construct the qubit-encoded projection image

$$IP_{ROI}(\theta, s),$$

which denotes the projection image generated from the ROI pixels represented by the binary decision variables. The second-stage optimization then seeks binary variables such that $IP_{ROI}(\theta, s)$ best matches the residual projection image $P_{res}(\theta, s)$.

*E. QUBO Formulation for ROI Refinement*

Let $\vec{q} \in \{0, 1\}^n$ denote the vector obtained by stacking all binary decision variables $q_k^{ij}$ associated with the ROI pixels. Since the qubit-encoded projection image is determined by these variables, we write it as $IP_{ROI}(\theta, s; \vec{q})$. The local refinement problem is formulated as

$$\min_{\vec{q} \in \{0,1\}^n} \sum_\theta \sum_s \left(P_{res}(\theta, s) - IP_{ROI}(\theta, s; \vec{q})\right)^2. \quad (5)$$

By substituting the difference-MAC-based pixel representation into $IP_{ROI}(\theta, s; \vec{q})$, the objective can be expanded into a quadratic polynomial over binary decision variables. Therefore, the problem can be written in standard QUBO form as

$$\min_{\vec{q} \in \{0,1\}^n} \vec{q}^\top Q \vec{q} + c, \quad (6)$$

where $Q$ is the QUBO matrix and $c$ is a constant term independent of $\vec{q}$. In particular, $c$ includes the $\sum_\theta \sum_s \left(P_{res}(\theta, s)\right)^2$ term arising from the quadratic expansion. This QUBO problem is solved using the D-Wave hybrid solver, and the optimized ROI is inserted into $I_{global}^{\setminus ROI}$ to obtain the final reconstruction.

*F. Evaluation metrics*

Since the main objective of the proposed framework is to improve reconstruction quality in the selected ROI, quantitative evaluation is primarily performed using ROI-based metrics. In this study, we report ROI-based error measures, specifically the root mean square error (RMSE) and the mean absolute error (MAE). Let $G^{ij}$ and $R^{ij}$ denote the ground-truth and reconstructed pixel values, respectively.

Let $\Omega_{ROI}$ denote the set of pixels inside the selected ROI, and let $|\Omega_{ROI}|$ denote the number of pixels in the ROI.
The ROI RMSE is defined as

$$RMSE_{ROI} = \sqrt{\frac{1}{|\Omega_{ROI}|} \sum_{(i,j)\in\Omega_{ROI}} (G^{ij} - R^{ij})^2}.$$

The ROI MAE is defined as

$$MAE_{ROI} = \frac{1}{|\Omega_{ROI}|} \sum_{(i,j)\in\Omega_{ROI}} |G^{ij} - R^{ij}|.$$

These metrics are used to evaluate how accurately the selected ROI is reconstructed under each pipeline. For coarse images, analogous RMSE and MAE values are computed over the non-ROI region in order to evaluate the background quality outside the selected local target.

## IV. Experimental Results

The proposed reconstruction framework was evaluated on three synthetic phantom samples with increasing structural complexity. Samples 1 and 2 were defined on $120 \times 120$ grids, whereas Sample 3 was defined on a $180 \times 180$ grid. In all cases, the selected ROI had size $60 \times 60$.

For Samples 1 and 2, all pipelines were tested under the same reduced-angle setting using 60 projection views throughout. In particular, QTR+QTR, FBP+QTR, and SART+QTR all used 60-view data for both the coarse reconstruction stage and the residual-based second-stage refinement. For Sample 3, the QTR+QTR pipeline remained fully in the 60-view setting, whereas the FBP+QTR and SART+QTR pipelines used 180-view data only for the coarse reconstruction stage; the second-stage residual based QTR refinement still used 60-view data. This additional setting was introduced because, when the Sample 3 FBP and SART coarse images were generated from only 60 projection views, the resulting ROI refinement also became spot-like and unstable.

Fig. 2 shows the final reconstruction results for the three phantom samples. Each row corresponds to one sample, and each column shows the ground truth, QTR+QTR, FBP+QTR, and SART+QTR results. For Samples 1 and 2, both QTR+QTR and SART+QTR recovered the ROI with high visual fidelity, whereas FBP+QTR showed more visible residual artifacts. For Sample 3, the difference among the pipelines became much larger. The QTR+QTR result preserved only the broad ROI structure and contained substantial spot-like artifacts, whereas the SART+QTR result reconstructed the ROI cleanly in the reported setting.

Fig. 3 presents the ROI absolute difference maps between the ground truth and each reconstruction result. For Samples 1 and 2, the difference maps for QTR+QTR and SART+QTR are both close to zero, while FBP+QTR retains more localized error. For Sample 3, the difference map of QTR+QTR contains widespread residual error, whereas the SART+QTR difference map is nearly zero.

Figure 4 shows CT images reconstructed using QCSTR instead of QTR, similar to the method used in Figure 2. The algorithms used for reconstructing each CT image can be seen in the figure.

These visual observations are consistent with the quantitative results in Table 1, which reports ROI RMSE and ROI MAE for all samples and reconstruction pipelines. Across all three samples, SART+QTR achieved the smallest average ROI error, with average ROI RMSE and ROI MAE of 0.0175 and 0.0005, respectively. For Samples 1 and 2, QTR+QTR also performed well, with ROI RMSE values of 0.0833 and 0.0707 and ROI MAE values of 0.0053 and 0.0044. In Sample 3, however, the QTR+QTR error increased substantially to 0.5367 in ROI RMSE and 0.2753 in ROI MAE, whereas SART+QTR achieved 0 for both metrics in the reported setting. The FBP+QTR pipeline showed larger ROI errors than SART+QTR in all three samples.

To compare the coarse-image behavior outside the ROI, Table 2 reports non-ROI RMSE and MAE for the coarse reconstructions. In these measurements, the FBP coarse images gave smaller average non-ROI error than the SART coarse images, with average non-ROI RMSE values of 0.1085 and 0.1326, respectively, and average non-ROI MAE values of 0.0644 and 0.0685. Thus, the ranking of the coarse-image non-ROI error did not match the ranking of the final ROI reconstruction error in Table 1.

Additional statistics of the coarse reconstructions showed that the FBP coarse images had larger negative minimum values than the corresponding SART coarse images in Samples 1 and 2. In particular, the minimum values of the FBP coarse reconstructions were approximately -0.45 and -0.47, whereas the corresponding SART values were about -0.07 in both samples.

TABLE I. ROI RECONSTRUCTION ERROR FOR ALL SAMPLES AND RECONSTRUCTION PIPELINES, REPORTED IN TERMS OF RMSE AND MAE. FOR SAMPLE 3, THE FBP+QTR AND SART+QTR PIPELINES USED 180-VIEW COARSE RECONSTRUCTION.

| Sample | Metric | Direct FBP | Direct SART | QTR→ QTR | FBP→ QTR | SART →QTR |
|---|---|---|---|---|---|---|
| 1 | RMSE | 0.0927 | 0.1745 | 0.0833 | 0.1780 | **0.0289** |
| | MAE | 0.0637 | 0.1215 | 0.0053 | 0.0317 | **0.0008** |
| 2 | RMSE | 0.0997 | 0.1816 | 0.0707 | 0.2749 | **0.0236** |
| | MAE | 0.0699 | 0.1279 | 0.0044 | 0.0756 | **0.0006** |
| 3 | RMSE | 0.0912 | 0.134 | 0.5367 | 0.4601 | **0.0000** |
| | MAE | 0.0509 | 0.0752 | 0.2753 | 0.2078 | **0.0000** |
| Average | RMSE | 0.0945 | 0.1634 | 0.2303 | 0.3043 | **0.0175** |
| | MAE | 0.0615 | 0.1082 | 0.0950 | 0.105 | **0.0005** |

TABLE II. NON-ROI ERROR OF THE COARSE IMAGES, REPORTED IN TERMS OF RMSE AND MAE, USED TO ANALYZE HOW THE COARSE-STAGE RECONSTRUCTION QUALITY AFFECTS THE SUBSEQUENT ROI REFINEMENT.

| Sample | Metric | QTR coarse | FBP coarse | SART coarse |
|---|---|---|---|---|
| 1 | RMSE | 0.2482 | **0.1229** | 0.1390 |
| | MAE | 0.1321 | 0.0742 | **0.0735** |
| 2 | RMSE | 0.2483 | **0.1226** | 0.1390 |
| | MAE | 0.1327 | 0.0746 | **0.0736** |
| 3 | RMSE | 0.2799 | **0.0801** | 0.1197 |
| | MAE | 0.1307 | **0.0445** | 0.0584 |
| Average | RMSE | 0.2588 | **0.1085** | 0.1326 |
| | MAE | 0.1318 | **0.0644** | 0.0685 |

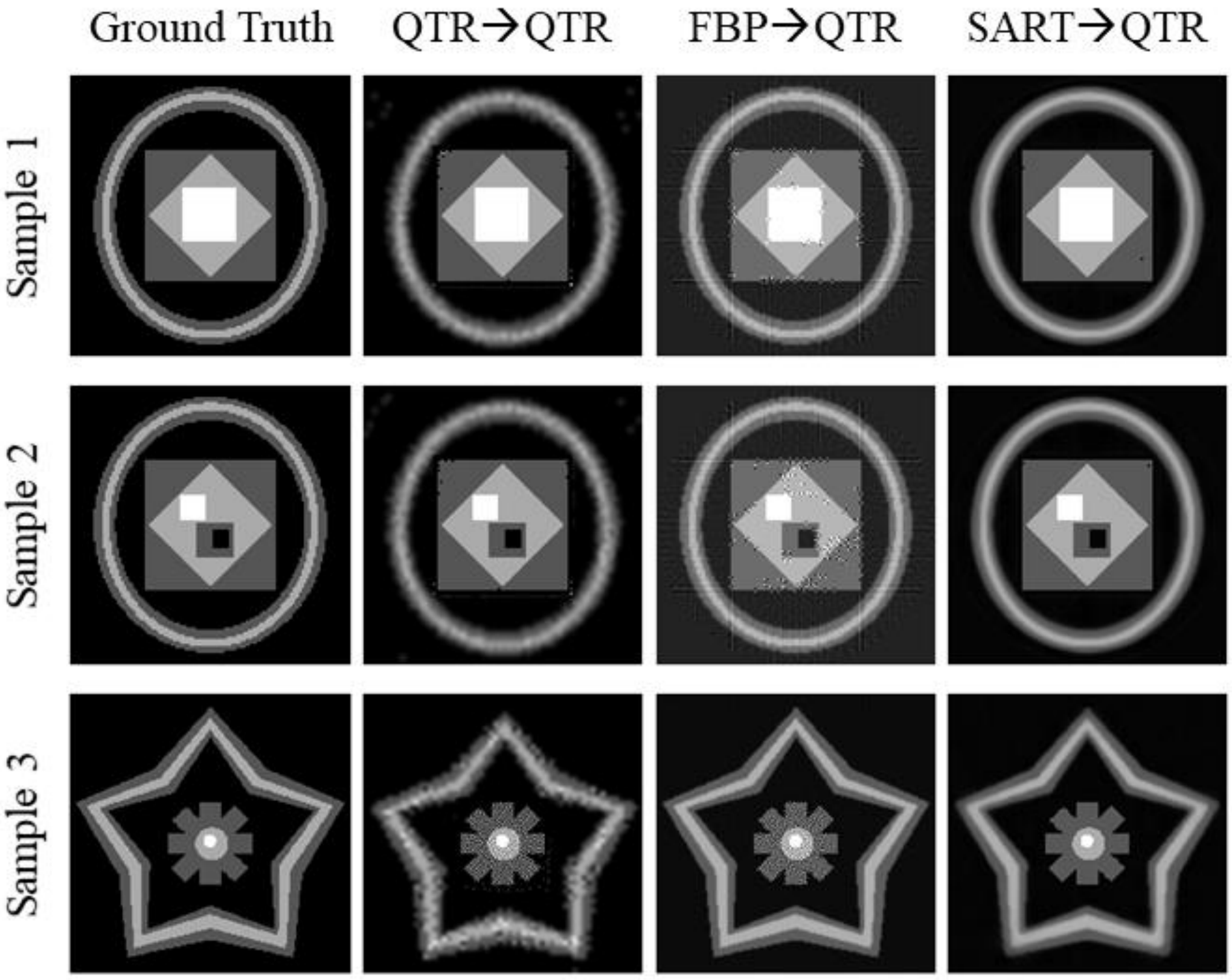

Fig. 3. Final reconstruction results for the three phantom samples. Samples 1 and 2 were defined on $\mathbf{120 \times 120}$ grids, whereas Sample 3 was defined on a $\mathbf{180 \times 180}$ grid; the ROI size was $\mathbf{60 \times 60}$ in all cases. Each row corresponds to one sample, and each column shows the ground truth, QTR+QTR, FBP+QTR, and SART+QTR results. For Sample 3, the FBP+QTR and SART+QTR results were obtained using a denser angular sampling in the coarse-reconstruction stage, while the second-stage ROI refinement was performed with the same residual-based QTR formulation.

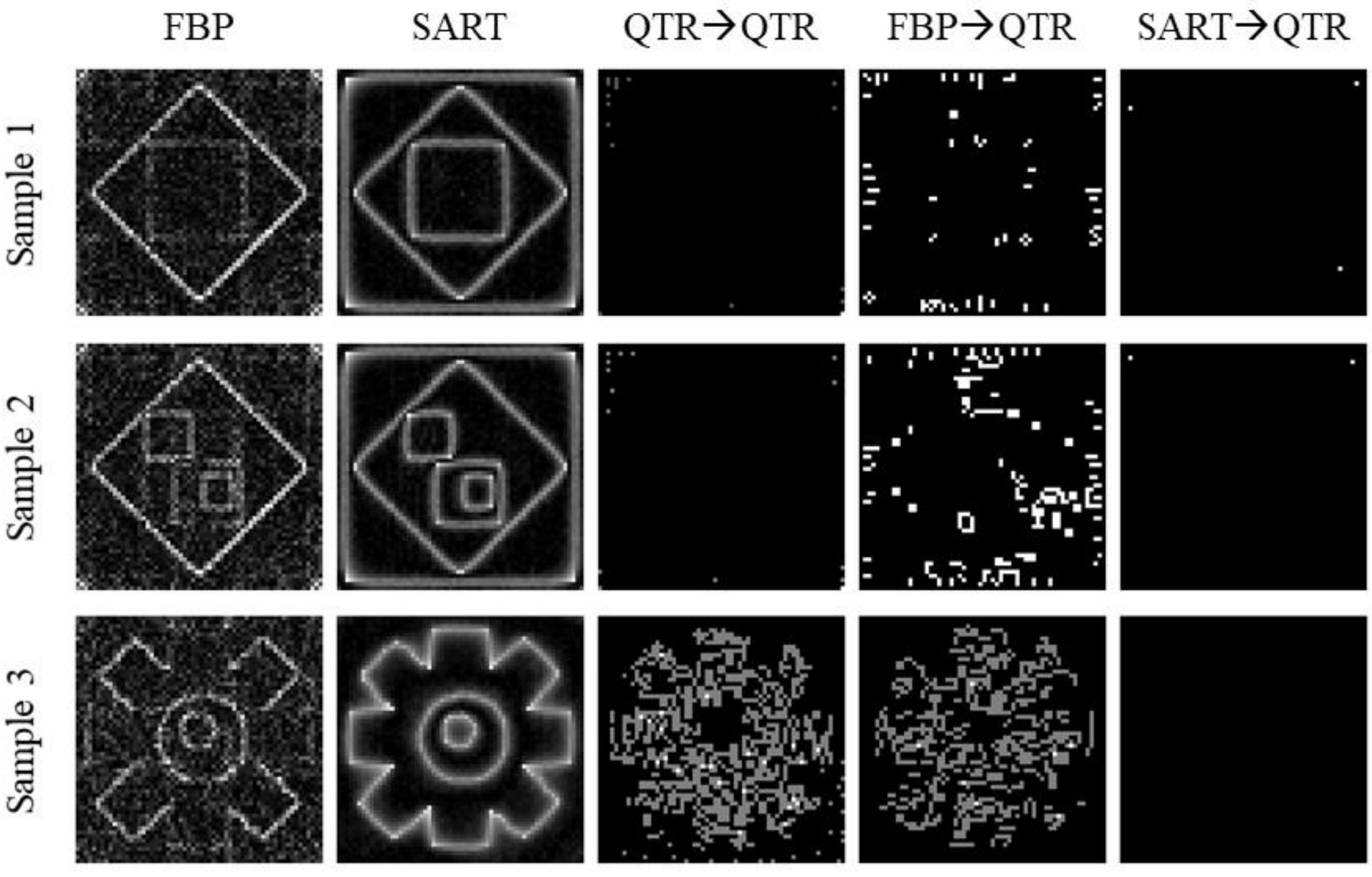

Fig. 2. ROI absolute difference maps $|\boldsymbol{GT - reconstruction}|$ for the three phantom samples. Samples 1 and 2 have size $\mathbf{120 \times 120}$, whereas Sample 3 has size $\mathbf{180 \times 180}$; the ROI size is $\mathbf{60 \times 60}$ in all cases. Each row corresponds to one sample, and each column corresponds to one reconstruction pipeline. Lower intensity indicates smaller reconstruction error inside the selected ROI. In QTR+QTR, both stages used 60 projection views, and the detector-position dimension was reduced to 60 before the first-stage QTR reconstruction. For Sample 3, the FBP+QTR and SART+QTR results used 180-view coarse reconstruction and 60-veiw second-stage QTR refinement.

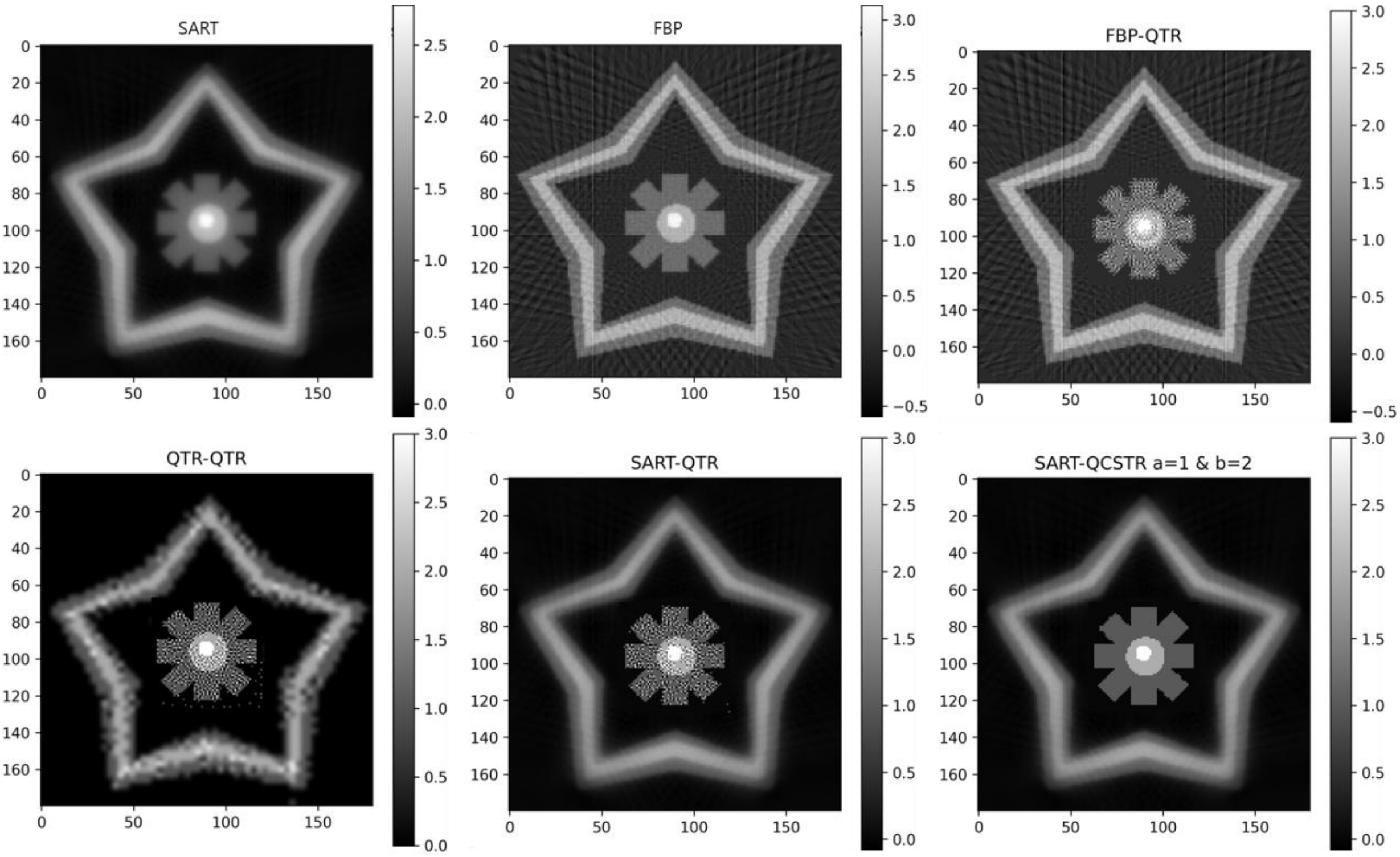

Fig.4. . Final reconstruction results for the phantom sample. Sample was defined on a $\mathbf{180} \times \mathbf{180}$ grid; the ROI size was $\mathbf{60} \times \mathbf{60}$ in all cases. Each row corresponds to one sample, and each column shows the ground truth, QCSTR+QCSTR, FBP+QCSTR, and SART+QCSTR results.

## V. Discussion

Mammography is the standard imaging modality for breast cancer screening; however, it has an important limitation. Its sensitivity can be substantially reduced in women with dense breasts, where lesions may be obscured by overlapping fibroglandular tissue. Fig. 5 presents an X-ray projection image of a composite sample consisting of a rod-shaped zirconia specimen with an artificially introduced crack embedded in an artificial sample characterized by a low X-ray MAC. Fig. 5a and 5b were acquired at projection angles differing by 90°. Cracks oriented horizontally are clearly visible in both projection images, whereas cracks with other orientations are difficult to identify and may be nearly indistinguishable at certain viewing angles. Despite the reconstruction accuracy of the QTR algorithm, high-resolution CT image reconstruction remains challenging because of current hardware limitations. In this paper, instead of reconstructing the entire sample, we propose a strategy for accurately reconstructing only selected regions of interest using the QTR algorithm and performing diagnostic imaging of multiple localized areas with a hybrid solver, analogous to targeted biopsy. This approach is expected to provide a practically useful framework for the medical imaging field, particularly in settings where rapid and focused image-based assessment is required.

The results show that the effectiveness of ROI-focused quantum refinement depends not only on the QUBO-based second-stage reconstruction itself, but also on the quality and stability of the coarse global estimate provided in the first stage. This dependence becomes particularly clear when comparing the behavior across the three phantom samples. For Samples 1 and 2, both QTR+QTR and SART+QTR achieved highly accurate ROI reconstruction, indicating that the fully quantum two-stage pipeline can remain competitive when the image size is moderate and the target structure is relatively manageable. In these cases, the reduced-angle setting with 60 projection views was sufficient for both the coarse stage and the refinement stage to produce near-exact ROI recovery.

However, the results for Sample 3 reveal a different regime. In this larger and more structurally demanding case, the QTR+QTR pipeline, which remained fully in the 60-view setting, no longer produced a stable high-fidelity ROI reconstruction. Instead, the reconstructed ROI retained only the broad target shape and showed strong spot-like artifacts. By contrast, when the coarse image for the FBP+QTR and SART+QTR pipelines was generated from denser angular sampling, the second-stage QTR refinement improved markedly, and SART+QTR achieved perfect reconstruction of the ROI in the reported setting. This contrast suggests that, as structural complexity increases, the role of the coarse stage shifts from being a simple initialization step to being a decisive component of the overall refinement process.

This observation supports an important interpretation of the proposed framework. The goal of the method is not necessarily to reconstruct the entire image quantum mechanically at every stage, but rather to allocate quantum optimization where it is most useful. In the present setting, this means using classical reconstruction to stabilize the global background and reserving QTR for the selected ROI. From this perspective, the strong performance of SART+QTR suggests that the practical advantage of the hybrid framework lies in separating global modeling from local quantum refinement, rather than forcing the entire reconstruction process into a uniformly reduced-angle or fully quantum setting.

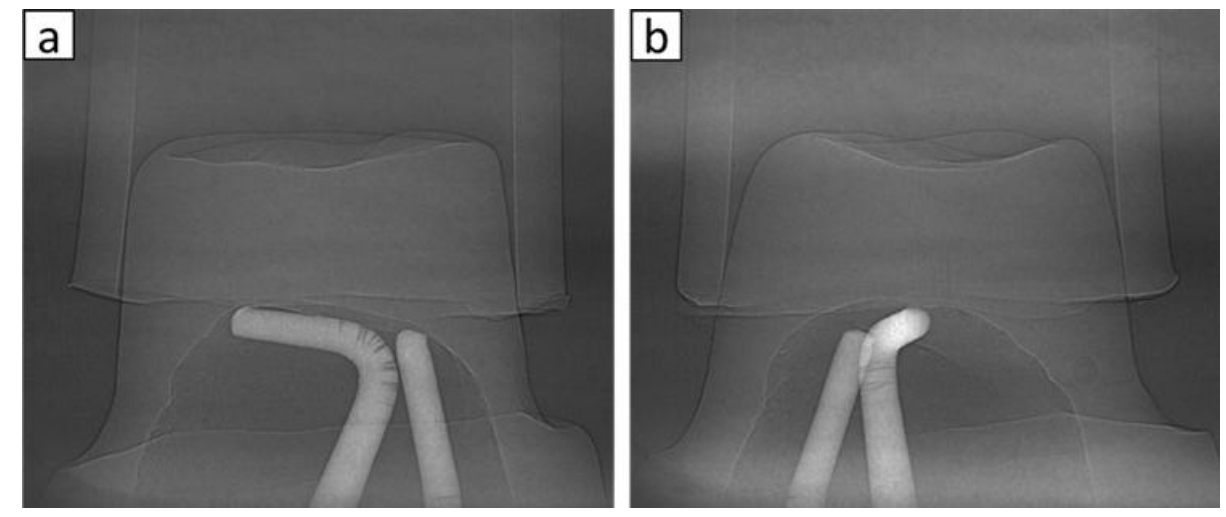

Fig. 5. X-ray projection images of a mixed sample composed of a high-X-ray-MAC inner specimen embedded within a low-X-ray-MAC outer sample. To investigate the detectability of an internal crack, an artificial crack was introduced into the inner specimen, and projection images were obtained at two angles differing by 90°: (a) and (b).

An interesting feature of the results is that the non-ROI coarse-image error does not directly predict the final ROI-refinement performance. In Table 2, the FBP coarse images often show smaller non-ROI RMSE and MAE than the corresponding SART coarse images. If average coarse-image error alone were the dominant factor, one might expect FBP+QTR to outperform SART+QTR. However, the opposite trend is observed in the final ROI reconstruction results. This indicates that the success of the second-stage QTR refinement is not determined solely by the magnitude of the coarse-image error outside the ROI.

A more plausible explanation is that the refinement stage is sensitive not only to average error, but also to the structural quality of the residual projection image used in the QUBO formulation. In the proposed method, the residual projection is defined as a signed quantity. Therefore, the refinement stage is influenced not only by how small the error is on average, but also by how coherently the coarse image explains the non-ROI contribution in the projection domain. In this sense, a coarse image with slightly larger average error may still be more favorable if it produces a cleaner and more structurally stable residual for the second-stage QTR optimization.

The out-of-range value statistics provide partial support for this interpretation. Although the proportion of out-of-range pixels was not dramatically different between FBP and SART, the FBP coarse images exhibited substantially larger negative excursions in Samples 1 and 2. In particular, the minimum values of the FBP coarse images were approximately $-0.45$ and $-0.47$, whereas the corresponding SART values were only about $-0.07$. Because the residual projection used in the QUBO formulation preserves sign information, these larger negative overshoots may distort the residual structure more strongly than is reflected by coarse-image RMSE or MAE alone. Thus, the signed residual appears to depend not only on average coarse-image fidelity but also on whether the coarse reconstruction remains physically and structurally well behaved.

At the same time, the additional diagnostic experiment with nonnegative FBP coarse reconstruction suggests that negative overshoot alone is not the whole explanation. When the negative values of the FBP coarse image in Sample 1 were clipped to zero before forming the residual and applying the second-stage QTR, the final ROI error improved slightly, but the overall reconstruction behavior did not change dramatically. This result is informative for two reasons. First, it indicates that negative overshoot does have some influence on the refinement process. Second, and more importantly, it suggests that the main difference between FBP+QTR and SART+QTR cannot be reduced to a simple range-correction issue. Instead, the broader structural quality of the coarse image—such as the smoothness, consistency, and projection-domain stability of the estimated background—appears to play a more important role.

This interpretation is also consistent with the visual results. In Samples 1 and 2, the QTR+QTR and SART+QTR pipelines both produced near-perfect ROI reconstructions, but SART+QTR still yielded the most spatially uniform difference maps. In Sample 3, where the image became larger and the structure more complex, the difference between these coarse-stage strategies became decisive. The fact that the SART-based coarse reconstruction led to the most stable final refinement suggests that coarse-stage regularity becomes increasingly important as the background becomes harder to approximate under limited angular sampling.

Taken together, the results suggest that the proposed framework should be understood as a hybrid strategy in which the first stage is responsible for providing a sufficiently stable global estimate and the second stage is responsible for allocating quantum optimization power to the ROI. In this sense, the main contribution of the method is not merely that it reconstructs a local region with fewer qubits, but that it shows how quantum refinement can be combined with classical reconstruction in a practically effective way. The improvement of SART+QTR over QTR+QTR in the most difficult sample indicates that reducing quantum burden in the coarse stage can actually improve the final reconstruction, provided that the remaining classical stage supplies a reliable global background.

The three samples also illustrate that the practical role of projection-angle reduction depends on structural complexity. In Samples 1 and 2, accurate ROI reconstruction was achieved while keeping both the coarse stage and the second-stage QTR refinement in the same 60-view reduced-angle setting, showing that the proposed framework can preserve the advantage of limited-angle acquisition in relatively moderate cases. In Sample 3, however, the results suggest that the more critical issue is not to reduce the number of projection views at every stage, but to secure a sufficiently stable global background estimate before ROI refinement. In this case, using denser angular sampling only for the coarse classical reconstruction while keeping the second-stage QTR refinement in the 60-view setting still allowed high-fidelity local reconstruction with limited quantum resources. This suggests that, for larger and more complex images, the most practical strategy is to stabilize the background classically and reserve quantum optimization for the selected ROI.

From a longer-term perspective, this ROI-prioritized strategy may also be relevant to future precision-focused imaging workflows in which attention is concentrated on a small number of diagnostically important local regions rather than distributed uniformly over the entire image, including settings where localized image-based assessment is more valuable than uniform full-field reconstruction. In such settings, selectively improving reconstruction fidelity in several suspicious subregions could be more valuable than enforcing uniformly high-fidelity reconstruction over the full field of view. Although the present study does not establish a direct clinical replacement for invasive procedures such as biopsy, it does suggest the feasibility of localized high-fidelity reconstruction as a potentially useful component in future diagnostic imaging strategies.

## VI. Conclusion

In this paper, we proposed a hybrid ROI-refinement framework for CT reconstruction under limited quantum resources. The method combines a coarse global reconstruction stage with a second-stage QTR refinement applied only to the selected ROI. This design makes it possible to reduce the burden on the quantum stage while still achieving high-fidelity reconstruction in the target region.

Experiments on three discrete phantom samples showed that the proposed framework can recover the ROI accurately across different structural settings. In the moderate-size cases, both QTR+QTR and SART+QTR produced highly accurate ROI reconstructions under the same 60-view reduced-angle setting. For the largest and most complex sample, however, the results showed that the quality of the coarse global estimate became decisive, and the best performance was obtained when the coarse background was stabilized classically and the quantum stage was reserved for ROI refinement. Among the tested pipelines, SART+QTR achieved the most stable overall performance.

These results suggest that the practical value of quantum-assisted CT reconstruction may lie less in replacing the entire reconstruction pipeline and more in identifying the stages where quantum optimization is most effective. In particular, the present results support a task-separated hybrid strategy in which classical reconstruction provides a reliable global background and quantum optimization is concentrated on the selected local region.

Future work will investigate how far the coarse-stage projection views can be reduced without destabilizing the final ROI refinement, and whether similar hybrid strategies remain effective for more realistic medical or industrial CT data.

## ACKNOWLEDGMENT

This research was supported by 'Creation of the quantum information science R&D ecosystem(based on human resources)' through the National Research Foundation of Korea(NRF) funded by the Korean government (RS-2023-NR057243). H. L. was supported by the National Research Foundation of Korea(NRF) grant funded by the Korea government(MSIT)(RS-2024-00352408).

## DATA AVAILABILITY

The synthetic phantom data used in this study were generated by the authors and are available from the corresponding author upon reasonable request.

## COMPETING FINANCIAL INTERESTS

The key idea developed in this paper were funded by QTomo Inc., which holds ownership of the patent for the algorithm.